\documentclass[preprint,showpacs,preprintnumbers,amsmath,amssymb,numerical]{revtex4}

\usepackage{graphicx}
\usepackage{dcolumn}
\usepackage{bm}
\usepackage{color}

\begin{document}


\title{Ultra low noise YBCO nanoSQUIDs implementing nanowires}


\author{R. Arpaia$^{1,2}$}

\author{M. Arzeo$^1$}

\author{S. Nawaz$^1$}

\author{S. Charpentier$^1$}

\author{F. Lombardi$^1$}

\author{T. Bauch$^1$}

\affiliation{$^1$ Quantum Device Physics Laboratory, Department of Microtechnology and Nanoscience, Chalmers University of Technology,  SE-41296 G\"{o}teborg, Sweden}

\affiliation{$^2$ CNR-SPIN, Dipartimento di Scienze Fisiche, Universit\`a degli Studi di Napoli ``Federico II'', I-80125 Napoli, Italy}

\date{\today}

\begin{abstract}
We present results on ultra low noise YBa$_2$Cu$_3$O$_{7-\delta}$ nano Superconducting QUantum Interference Devices (nanoSQUIDs). To realize such devices, we implemented high quality YBCO nanowires, working as weak links between two electrodes.  We observe critical current modulation as a function of an externally applied magnetic field in the full temperature range below the transition temperature $T_C$. The white flux noise below 1 $\mu \Phi_0/\sqrt{\mathrm{Hz}}$ at $T = 8$~K makes our nanoSQUIDs very attractive for the detection of small spin systems.
\end{abstract}

\maketitle

The development of quantum limited magnetic flux sensors has recently gained a lot of attention for the possibility to detect the magnetic moment of nanoscaled systems, with the ultimate goal of the observation of a single spin. Such sensors are of fundamental importance for applications ranging from spintronics and spin-based quantum information processing to fundamental studies of nano-magnetism in molecules and magnetic nano-clusters.
A nano-scale Superconducting QUantum Interference Device (nanoSQUID) is indeed a promising candidate to reach this ambitious goal \cite{Lam, Hao, Gallop}. A SQUID loop on the nanometer scale is a crucial requirement to achieve the necessary flux sensitivity and spacial resolution \cite{Foley}. 

The downscaling of tunnel junction based SQUIDs is an extremely challenging task \cite{NanoDavid, NatureNavid}. In particular, scaling down the dimensions of a conventional tunnel junction to nanometer size implies several drawbacks such as the deterioration of the tunnel barrier, with increased critical current/resistance noise \cite{David}, and small critical current values, limiting the working operation range of the SQUIDs far below the transition temperature of the superconducting material used. For these reasons during the recent years a lot of effort has been put into the development of nanoSQUIDs implementing superconducting nanowires in a Dayem bridge configuration \cite{Bezryadin, carmine}. At the moment, the realization of such nanoSQUIDs is well established for Low critical Temperature Superconductors (LTS) \cite{siddiqi}. NanoSQUIDs made of High critical Temperature Superconductors (HTS) might extend the operational working temperature (from mK to above 77 K) and the range of  magnetic fields that can be applied to manipulate spins compared to Nb based nanoSQUIDs. 

Several attempts to fabricate HTS nanoSQUIDs, implementing YBCO Dayem bridges, have been made during the last few decades \cite{Schneider, Pedyash, Wu}. However a proper SQUID behavior, with a periodic modulation of the critical current in the full temperature range below $T_C$ has never been observed. These results suggest a severe degradation of the YBCO nanostructures during fabrication, occurring because of chemical instability of this material and high sensitivity to defects and disorder due to the very short coherence length $\xi$.

In this letter, we present measurements on YBCO nanoSQUIDs, realized with Dayem bridges with cross sections down to 50x50 nm$^2$. In contrast to previous works \cite{Schneider, Pedyash, Wu} our nanoSQUIDs show critical current modulations as a function of an externally applied magnetic flux in the full temperature range below the transition temperature, $T_C$, of the devices. Both the modulation depth and the period in magnetic field are in good quantitative agreement with numerical computations. Moreover, the ultra low white flux noise below 1 $\mu \Phi_0/$Hz$^{1/2}$, that we have measured  above 10 kHz, makes these devices appealing for the investigation of small spin systems.

The Dayem bridges are realized by using YBCO nanowires fabricated using an improved nanopatterning procedure \cite{shahidPRL, arpaiaIEEE, shahidPhC}. The high value of the critical current achieved in our nanostructures demonstrate that the superconducting properties close to the as grown films are preserved. As a consequence, these nanostructures represent also model systems to investigate the instrinsic properties of HTS, for instance to study the fluxoid quantization in superconducting loops \cite{Sochnikov, Carillo}. 

\begin{figure}[tb!]
\begin{center}
\includegraphics[height=8cm]{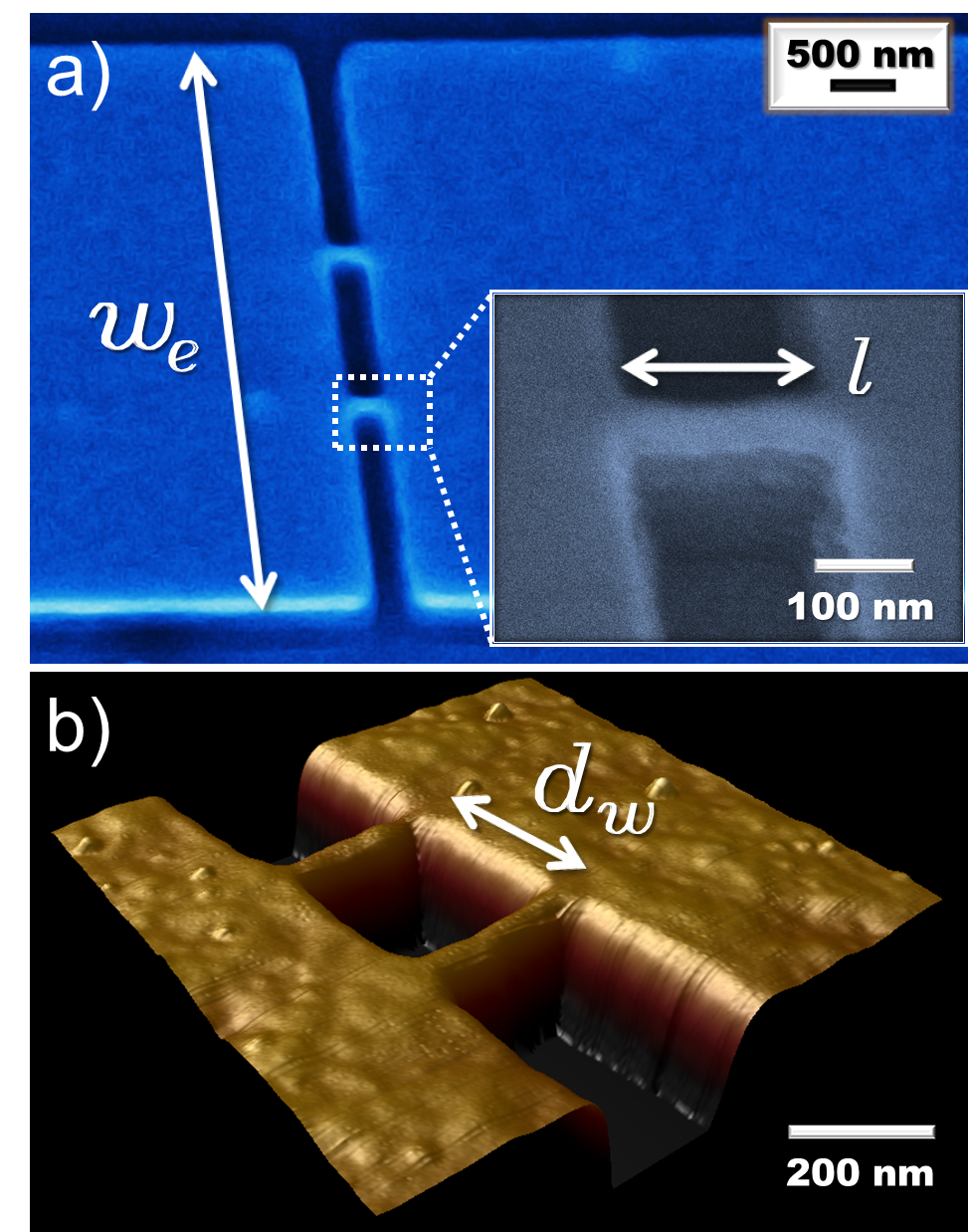}
\caption{a) SEM and b) AFM pictures of two nanoSQUIDs in the Dayem bridges configuration. The loop areas, respectively of 200x970 nm$^2$ and 200x150 nm$^2$  are realized with two parallel YBCO nanowires of length $l$, capped with Au and placed at a distance $d_w$, connecting two wider electrodes with width $w_e$.}
\label{fig:Fig1}
\end{center}
\end{figure}

A 50 nm thick YBCO film is deposited on a (110) MgO substrate by Pulsed Laser Deposition (PLD). The film has a very sharp transition with an onset at $T_C = 85$ K. For comparison, we have also patterned commercial YBCO films grown on (001) MgO substrates, provided by Theva GmbH, with a $T_C$ onset of 86 K. The nanostructures are defined by an e-beam lithography defined carbon mask and a very gentle Ar$^+$ ion milling. The nanopatterning procedure is described in detail in Refs. \citenum{shahidPRL, arpaiaIEEE, shahidPhC}. Fig.  \ref{fig:Fig1} shows images of typical nanoSQUIDs consisting of two nanowires in parallel, whose length $l$ is in the range 100 - 200 nm, connecting two wide electrodes with a width $w_e$  of nominally 4 $\mu$m. Different loop areas have been achieved, by varying the distance $d_w$ between the wires in the range 100 - 1000 nm. All the dimensions have been confirmed by scanning electron microscopy (SEM).
 
Electrical transport properties of the devices have been performed in a $^3$He cryostat. The current voltage characteristics ($IVC$s) were recorded using a 4-point measurement scheme. The nanoSQUIDs have a critical temperature very close to that of the bare films (differing not more than 1 K) and very high critical current densities $J_C$ at 300 mK: on the devices patterned on (001) MgO the average $J_C$ values per wire are in the range $7 - 9 \cdot 10^7$ A/cm$^2$; on devices on (110) MgO they are of the same order of magnitude, though slightly lower. \cite{note}

\begin{figure}[tb!]
\begin{center}
\includegraphics[height=8cm]{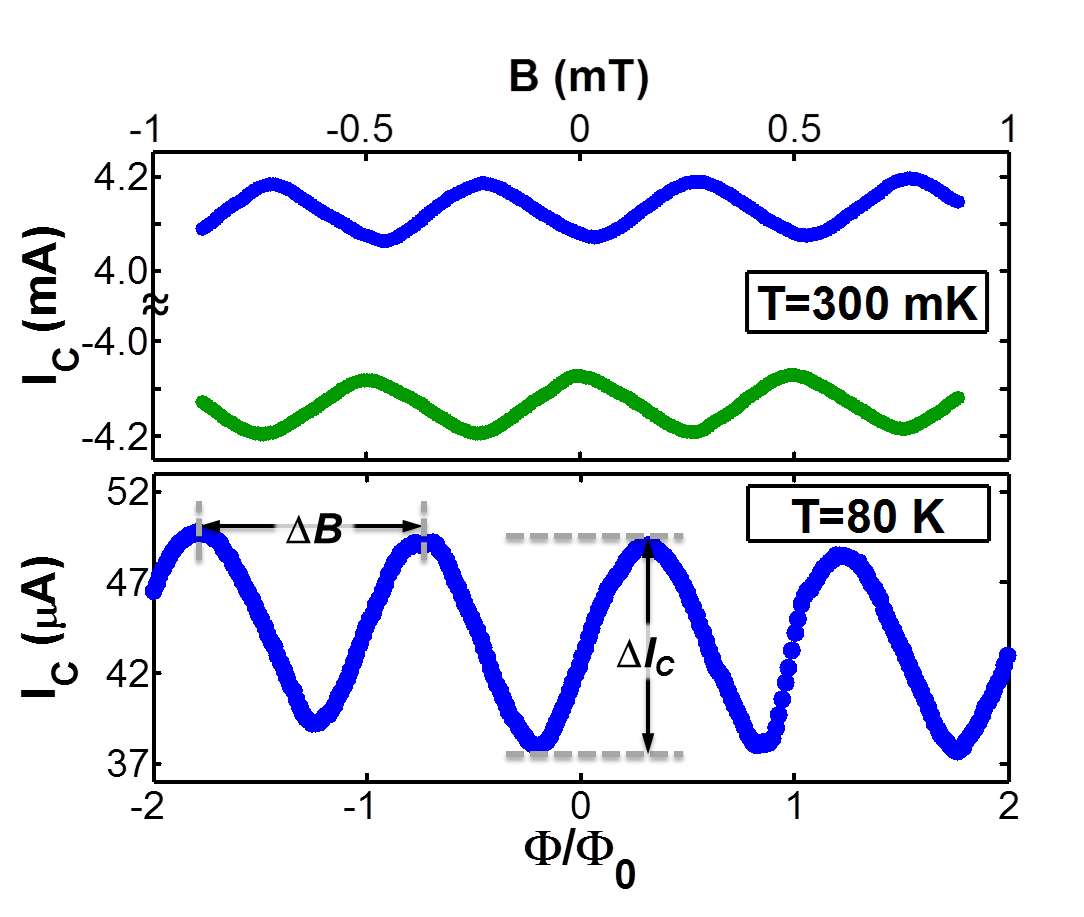}
\caption{Critical current as a function of the applied magnetic field measured on the same device at $T = 300$ mK ({\itshape upper panel}) and close to the $T_C$ ({\itshape lower panel}). The nanoSQUID, with a loop area of 130x1000 nm$^2$, is patterned on a (001) MgO substrate.}
\label{fig:Fig2}
\end{center}
\end{figure}

In Fig. \ref{fig:Fig2} we show the critical current of a nanoSQUID as a function of an externally applied magnetic field.  Modulations of the critical current have been observed in the whole temperature range, up to the critical temperature of the devices. Here, the critical current has been measured by ramping the current and detecting when the voltage exceeded a voltage criterium, the latter being determined by the noise level and the shape of the $IVC$ (usually a value of $\sim$ 2 $\mu$V has been considered). From the critical current modulation we extract the modulation period $\Delta B$ and the relative critical current modulation depth $\Delta I_C/I_C^{max}$, with $\Delta I_C$  being the difference between the maximum $I_C^{max}$ and the minimum $I_C^{min}$ values of the critical current.  

To calculate numerically the expected $\Delta I_C$, we have followed the approach by Tesche and Clarke \cite{Tesche}. For this purpose, the knowledge of the current-phase relation (CPR) of the bridges and the inductance of the electrodes is required. Concerning the CPR, our bridges are long nanowires, $l\!\gg\!\xi$ ($\xi \sim 2$ nm is the YBCO coherence length in the a-b plane), with cross section $wt\!\ll\!{\lambda}^2$ ($w$ and $t$ are respectively the width and the thickness of the nanowires, while $\lambda$ is the London penetration depth in the a-b plane). In this limit, the CPR is given by the Likharev and Yakobson expression \cite{LikYak, Likharev, Kupriyanov}
\begin{equation} \label{eq: LY}
J_s = \frac{{\Phi}_0}{2\pi{\mu}_0 \xi {\lambda}^2}\left[\left(\frac{\xi}{l}\right)\phi - {\left(\frac{\xi}{l}\right)}^3 {\phi}^3\right]\; ,
\end{equation}
where $J_s \!=\! I/wt$ is the superconducting current density, $\Phi_0=h/2e$ is the flux quantum, ${\mu}_0$ is the vacu\-um permeability and $\phi$ is the phase difference between the two ends of the wire. In case the critical current is limited by phase slips, the maximum phase difference is given by $\phi_d = l/\sqrt{3} \xi$. However for bridges wider than $4.4\xi$ the critical current is reached once vortices can overcome the bridge edge barrier. This occurs for a phase difference $\phi_v = l/2.718 \xi \simeq 0.64 \phi_d$ \cite{Bulaevskii}.
For $|\phi|<\phi_v$ the expression of the CPR (eq.(\ref{eq: LY})) can be reasonably approximated by the linear term:
\begin{equation} \label{eq: LY_approx}
I = \frac{{\Phi}_0}{2\pi L _k} \phi \; ,
\end{equation}
where $L_k$ is the kinetic inductance of the wire, given by $({\mu}_0 {\lambda}^2 l)/(w t)$. Each nanowire inside the loop behaves therefore as an inductor, where the phase difference between the two ends grows linearly with the bias current.  Indeed, the inductance of a wire with cross section $wt\!\!\ll\!\!{\lambda}^2$ is dominated by the kinetic inductance with a negligible contribution of the geometric inductance $L_g\! \simeq \!{\mu}_0 l$.
 
\begin{figure}[b]
\begin{center}
\includegraphics[height=8cm]{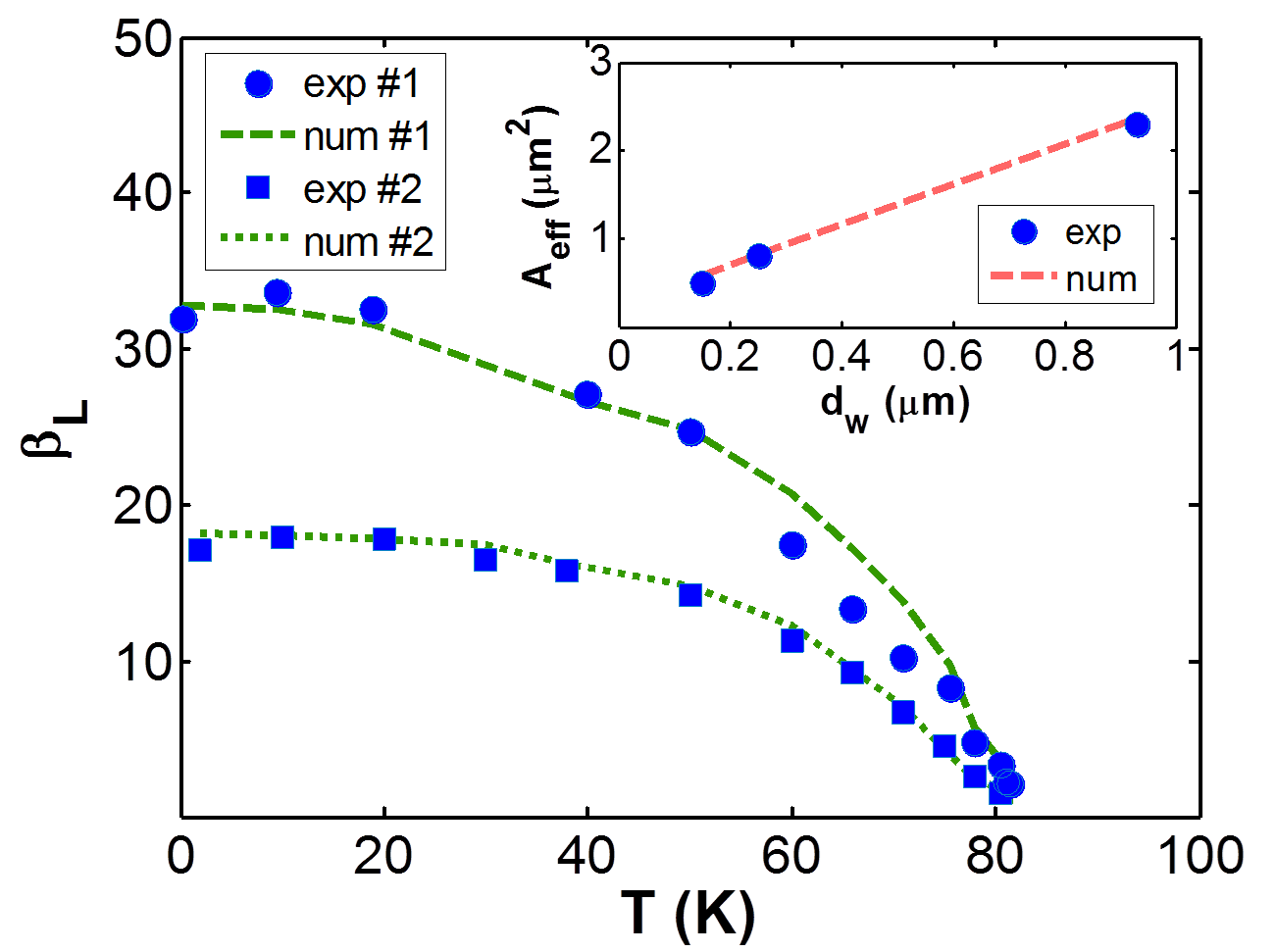}
\caption{Comparison between the experimental (dots) and the theoretical (lines) values of the screening inductance factor $\beta_L$ as a function of the temperature for nanoSQUIDs patterned both on (001) MgO substrate (\#1, whose IVCs are shown in Fig.\ref{fig:Fig2}) and (110) MgO substrate (\#2). The experimental values of $\beta_L(T)$ have been extracted from the critical current modulation depths as $I_C^{max}/\Delta I_C$, while the theoretical ones have been obtained from the definition of $\beta_L(T)$, determining the loop inductance through numerical computation. ({\itshape inset}) Comparison between the experimental (dots) and the calculated (lines) values of the effective area $A_{eff}$ for several devices patterned on (110) MgO, with same electrodes width (4 $\mu$m) and wires length (150 nm) but different distance $d_w$ between the wires.} 
\label{fig:Fig3}
\end{center}
\end{figure}

From numerical calculations of the current modulation using the CPR of eq.(\ref{eq: LY_approx}) we obtain that \cite{Faucher, Podd}
\begin{equation} \label{eq: betaL}
\frac{\Delta I_C}{I^{max}_C}=  \frac{1}{\beta_L} \; ,
\end{equation}
where $\beta_L=I_C^{max}L_{loop}/\Phi_0$ is the screening inductance factor. Here, $L_{loop}$ is the total inductance of the SQUID loop, including the contributions both from the electrodes and from the wires.  This scaling behavior is also observed for SQUIDs containing Josephson junctions with sinusoidal CPR in the limit  $I_C^{max}L_{loop}\! \gg\! \Phi_0$ \cite{Tesche}. Since we can neglect the nonlinearity of the current-phase relation (eq.(\ref{eq: LY})),  the total inductance of our SQUID loop can be calculated from the Maxwell and London equations describing the Meissner state \cite{Khapaev, Jesper, shahidPRL}.  As expected from their dimensions \cite{Hasselbach}, our devices are governed by the kinetic inductance $L_k(T)$ of the nanowires ($\sim 15$ pH at 300 mK, one order of magnitude higher than the geometrical value). For the temperature dependence of the loop inductance we use the two-fluid model for the London penetration depth: $\lambda (T)=\lambda_0 [1-{(T/T_C)}^2]^{-1/2}$ \cite{Tinkham}, with $\lambda_0$ value of the London depth at zero temperature.  The numerically calculated loop inductance $L_{loop}^{num}(T)$ allows to determine $\beta_L^{num}(T)=I_C^{max}(T)L_{loop}^{num}(T)/\Phi_0$ (here, the $I_C^{max}(T)$ values are those extracted from the measurements). We can now fit the experimentally determined parameter,  $\beta_L^{exp}$, defined through eq.(\ref{eq: betaL}) as $\beta_L^{exp}(T)\! =  \! I_C^{max}(T)/\Delta I_C(T)$ (see solid symbols in Fig.\ref{fig:Fig3}) with the numerically calculated temperature dependent $\beta_L^{num}$, using $\lambda_0$ as the only fitting parameter. As shown in Fig. \ref{fig:Fig3}, the agreement between data and numerical calculations is very good using $\lambda_0 = 260$ nm (which is a typical value for thin YBCO films \cite{Zaitsev}), in the whole temperature range and for all the measured devices, both fabricated on (110) and (001) MgO. In particular, when the temperature increases, the critical current modulation depth becomes bigger as a consequence of the reduction of the critical current $I_C^{max}$: both $\beta_L^{exp}$ and $\beta_L^{num}$ decrease, approaching to 1 when the temperature is close to $T_C$.

We now focus on the periodicity $\Delta B$ of the critical current modulations. In the inset of Fig. \ref{fig:Fig3} we show the experimentally determined effective area $A_{eff}^{exp}\!=\!\Phi_0/\Delta B$ of nanoSQUIDs having different distances $d_w$ between the nanowires. These effective areas $A_{eff}^{exp}$ are far larger than the geometrical areas $A_g\!=\!d_w\!\cdot l$, defined by the distance and the length of the two wires. This can be understood considering that the superconducting phase gradient induced in the wide electrodes by the screening currents contributes to the total phase difference between the two wires resulting in an effective area which is larger than the geometric loop area \cite{Rosenthal}. These experimentally determined values of $A_{eff}^{exp}$  have been compared with those, calculated numerically, following Ref. \citenum{ClemPRB}, $A_{eff}^{num}\!=\!m/I_{cir}$. Here, $m\!= \! \frac{1}{2} \int \vec r \times \vec j d\vec r$ is the magnetic moment generated by a circulating current $I_{cir}$ around the SQUID loop. The result, presented in the inset of Fig. \ref{fig:Fig3}, shows a good agreement between theoretically and experimentally determined values of the effective area. In particular, our calculations show that the effective area is proportional to the product of the wire distance $d_w$ and the electrode width $w_e$, $A_{eff} \! \propto \! w_e \cdot d_w$. A similar dependency has been analytically found  for the effective area in Ref. \citenum{Bezryadin}, in the limit $d_w\!\ll\! w_e$ and $w_e\!\ll\! {\lambda}^2/t$. 

\begin{figure}[t]
\begin{center}
\includegraphics[height=7cm]{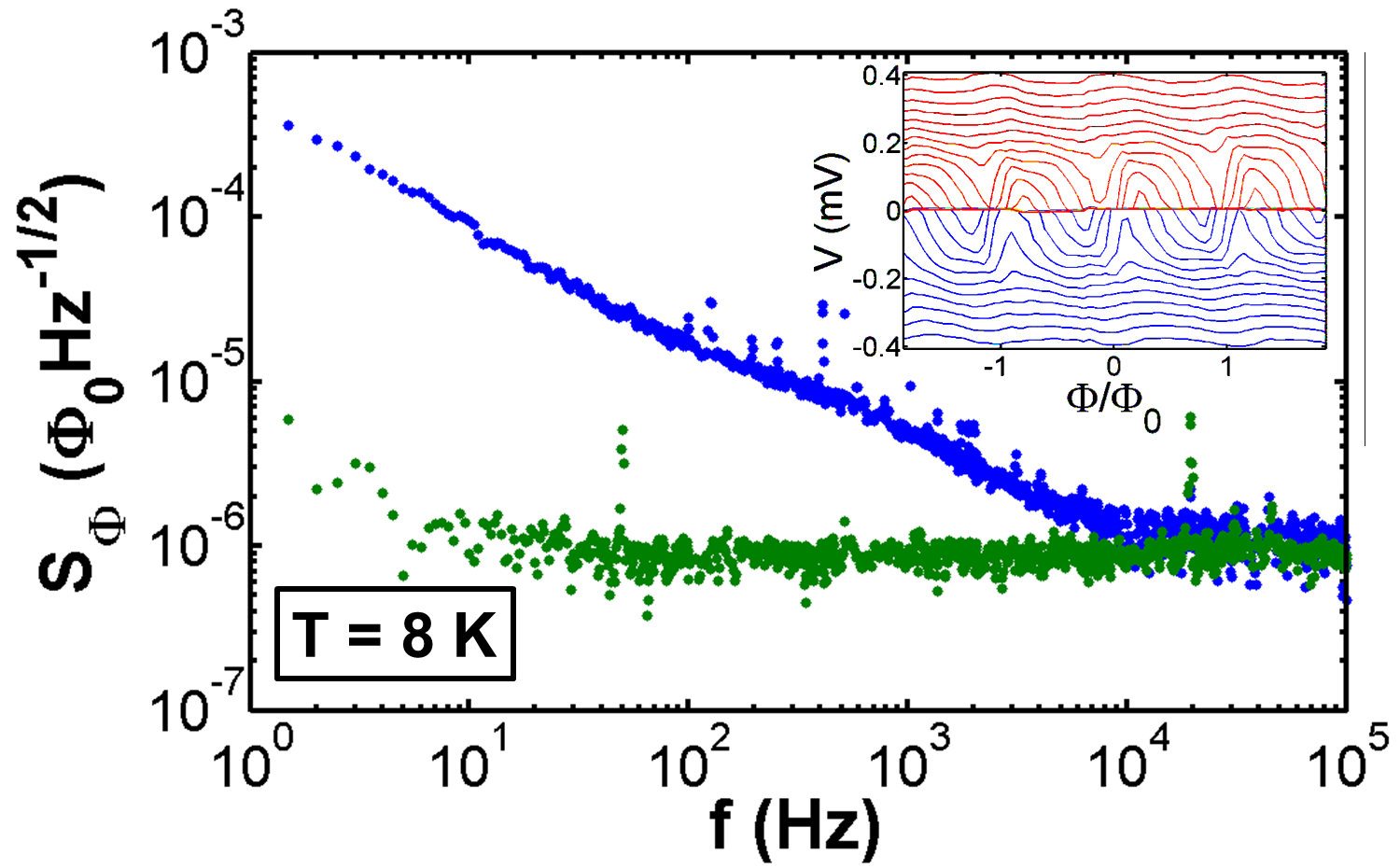}
\caption{Flux noise spectral density $S_{\Phi}$ vs frequency $f$, measured at $T = 8$ K on a nanoSQUID grown on (110) MgO and with a geometrical loop area $A_g = 0.1$ $\mu$m$^2$. Green dots are the amplifier background noise, blue dots represent the sum of the nanoSQUID and of the amplifier noise. In the inset,  $V(\Phi)$ of the device are shown for $I=[-2.5, 2.5]$ mA (in 26 $\mu$A steps), among which a voltage modulation of 0.2 mV (peak-to-peak) is present, corresponding to a transfer function $V_{\Phi}= 1.5$ mV/$\Phi_0$.}
\label{fig:Fig4}
\end{center}
\end{figure}

We have measured the flux noise of a nanoSQUID at a bias current slightly above the critical current and at a magnetic flux bias where the slope of the voltage modulations $V(\Phi)$ (see inset of Fig. \ref{fig:Fig4}) is maximized $V_{\Phi}=$ max$(\left | \partial V/\partial \Phi \right |)$. Using a cross correlation measurement scheme \cite{Sampietro} we achieved an amplifier input white noise level of  $\simeq 1.5$ nV/$\sqrt{\mathrm{Hz}}$, which includes also the thermal noise of the resistive voltage lines connecting the nanoSQUID to the amplifiers. From the measured voltage noise density $S_v$ we can calculate the flux noise density $S_\Phi = S_v/V_\Phi$. In Fig. 4 we show the magnetic flux noise measured on a nanoSQUID at $T = 8$ K. Above 10 kHz the measured white flux noise is $S_{\Phi}=1.2$ $\mu\Phi_0/\sqrt{\mathrm{Hz}}$, which is the sum of the intrinsic nanoSQUID flux noise and the noise added from the amplifier. From the measured value of the amplifier noise (see Fig. \ref{fig:Fig4}), $S_{\Phi}^{a} \simeq 1$ $\mu\Phi_0/\sqrt{\mathrm{Hz}}$ we can determine the upper limit for the intrinsic flux noise of the nanoSQUID: $ S_{\Phi}^{nS} =\sqrt{{S_{\Phi}}^2-{S_{\Phi}^{a}}^2} \simeq 0.7$ $\mu\Phi_0/\sqrt{\mathrm{Hz}}$. This is among the lowest values for YBCO nanoSQUIDs reported in literature \cite{Schwarz}, corresponding to a predicted spin sensitivity of only 50 $\mu_B$ per $\sqrt{\mathrm{Hz}}$, where $\mu_B$ is the Bohr magneton \cite{Nagel}. 
At frequencies below 10 kHz the noise spectrum is dominated by 1/$f$ noise. Since the measured 1/$f$ voltage noise spectra do not depend on the flux bias (data not shown), we attribute the 1/$f$ spectrum to critical current noise. The study of the origin of critical current noise in YBCO nanobridges and the implementation of a bias reversal SQUID readout electronics to minimize the effect of critical current noise on the measured flux noise \cite{Drung} will be subject of future work. 

In conclusion, we have fabricated YBCO nanoSQUIDs, realized in Dayem bridge configuration, working in the full temperature range. The high quality of the nanowires embedded in the loop is proved by the high critical currents they carry and by the observation of critical current modulations as a function of an externally applied magnetic field in the entire temperature range up to $T_C$ ($\sim 83$~K). Both the depth and periodicity of the measured modulations are in good agreement with numerical calculations, showing that the loop inductance is dominated by the kinetic inductance of the wires and the effective area is strongly affected by the screening currents induced in the electrodes. Finally, our devices exhibit an extremely low white flux noise above 10 kHz below 1 $\mu\Phi_0/\sqrt{\mathrm{Hz}}$, making them very attractive for many applications, as for the investigation of the magnetic moment in small ensembles of spins in a wide range of temperatures and magnetic fields. At the same time the study of the fluxoid quantization in these nanoSQUID loops, preserving pristine superconducting properties, close to the as grown films, could shed light on the microscopic mechanism leading to high critical temperature superconductivity.


\begin{thebibliography}{40}
\bibitem{Lam} S. K. H. Lam and D. L. Tilbrook,  \href{http://dx.doi.org/10.1063/1.1554770}{Appl. Phys. Lett.}, {\bf{82}}, 1078 (2003).
\bibitem{Hao} L. Hao, J. C. Macfarlane, J. C. Gallop, D. Cox, J. Beyer, D. Drung, and T. Schurig,  \href{http://dx.doi.org/10.1063/1.2917580}{Appl. Phys. Lett.}, {\bf{92}}, 192507 (2008).
\bibitem{Gallop} J. Gallop,  \href{http://dx.doi.org/10.1088/0953-2048/16/12/055}{Supercond. Sci. Technol.}, {\bf{16}}, 1575 (2003).
\bibitem{Foley} C. P. Foley and H. Hilgenkamp,  \href{http://dx.doi.org/10.1088/0953-2048/22/6/064001}{Supercond. Sci. Technol.}, {\bf{22}}, 064001 (2009).
\bibitem{NanoDavid} D. Gustafsson, H. Pettersson, B. Iandolo, E. Olsson, T. Bauch, and F. Lombardi, \href{http://dx.doi.org/10.1021/nl103311a}{Nano Lett.}, {\bf{10}}, 4824 (2010).
\bibitem{NatureNavid} D. Gustafsson, D. Golubev, M. Fogelstr\"{o}m, T. Claeson, S. Kubatkin, T. Bauch, and F. Lombardi, \href{http://dx.doi.org/10.1038/nnano.2012.214}{Nat. Nanotechnol. }, {\bf{8}}, 25 (2013).
\bibitem{David} D. Gustafsson, F. Lombardi, and T. Bauch,  \href{http://dx.doi.org/10.1103/PhysRevB.84.184526}{Phys. Rev. B}, {\bf{84}}, 184526 (2011).
\bibitem{Bezryadin} D. S. Hopkins, D. Pekker, P. M. Goldbart, and A. Bezryadin, \href{http://dx.doi.org/10.1126/science.1111307}{Science}, {\bf{308}}, 1762 (2005).
\bibitem{carmine}  C. Granata, E. Esposito, A. Vettoliere, L. Petti, and M Russo,  \href{http://dx.doi.org/10.1088/0957-4484/19/27/275501}{Nanotechnology}, {\bf{19}}, 275501 (2008).
\bibitem{siddiqi}  R. Vijay, E. M. Levenson-Falk, D. H. Slichter, and I. Siddiqi,  \href{http://dx.doi.org/10.1063/1.3443716}{Appl. Phys. Lett.}, {\bf{96}}, 223112 (2010).
\bibitem{Schneider} J. Schneider, M. M\"{u}ck, and R. W\"{o}rdenweber,  \href{http://dx.doi.org/10.1063/1.113037}{Appl. Phys. Lett.}, {\bf{65}}, 2475 (1994).
\bibitem{Pedyash}  M. V. Pedyash, D. H. A. Blank, and H. Rogalla,  \href{http://dx.doi.org/10.1063/1.115708}{Appl. Phys. Lett.}, {\bf{68}}, 1156 (1996).
\bibitem{Wu} C. H. Wu, Y. T. Chou, W. C. Kuo, J. H. Chen, L. M. Wang, J. C. Chen, K. L. Chen, U. C. Sou, H. C. Yang, and J. T. Jeng, \href{http://dx.doi.org/10.1088/0957-4484/19/31/315304}{Nanotechnology}, {\bf{19}}, 315304 (2008).
\bibitem{shahidPRL} S. Nawaz, R. Arpaia, F. Lombardi, and T. Bauch,  \href{http://dx.doi.org/10.1103/PhysRevLett.110.167004}{Phys. Rev. Lett.}, {\bf{110}}, 167004 (2013).
\bibitem{arpaiaIEEE} R. Arpaia, S. Nawaz, F. Lombardi, and T. Bauch, \href{http://dx.doi.org/10.1109/TASC.2013.2247454}{IEEE Trans. Appl. Supercond.}, {\bf{23}}, 3 (2013).
\bibitem{shahidPhC} S. Nawaz, R. Arpaia, T. Bauch, and F. Lombardi,  \href{http://dx.doi.org/10.1016/j.physc.2013.07.011}{Physica C}, {\bf{495}}, 33 (2013).
\bibitem{Sochnikov} I. Sochnikov, A. Shaulov, Y. Yeshurun, G. Logvenov, and I. Bozovic, \href{http://dx.doi.org/10.1038/NNANO.2010.111}{Nat. Nanotechnol.}, {\bf{5}}, 516 (2010).
\bibitem{Carillo} F. Carillo, G. Papari, D. Stornaiuolo, D. Born, D. Montemurro, P. Pingue, F. Beltram, and F. Tafuri,  \href{http://dx.doi.org/10.1103/PhysRevB.81.054505}{Phys. Rev. B}, {\bf{81}}, 054505 (2010).
\bibitem{note} To better define squared loop geometries even for very small areas, we exposed wider electrodes and narrower wires in two different e-beam lithographic steps: as a consequence, we observed that only films grown on (110) MgO are very sensitive to the consequent double baking treatment.
\bibitem{Tesche} C. D. Tesche and J. Clarke,  \href{http://dx.doi.org/10.1007/BF00655097}{J. Low Temp. Phys.}, {\bf{29}}, 301 (1977).
\bibitem{LikYak} K. K. Likharev and L. A. Yakobson,  \href{}{Sov. Phys. Tech. Phys.}, {\bf{20}}, 950 (1976).
\bibitem{Likharev} K. K. Likharev,  \href{http://dx.doi.org/10.1103/RevModPhys.51.101}{Rev. Mod. Phys.}, {\bf{51}}, 101 (1979).
\bibitem{Kupriyanov} M. Yu. Kupriyanov and K. K. Likharev, Fiz. Tverd. Tela {\bf{16}}, 2829 (1974) [Sov. Phys. Solid State {\bf{16}}, 1835 (1975)]. 
\bibitem{Bulaevskii} L. N. Bulaevskii, M. J. Graf, C. D. Batista, and V. G. Kogan, \href{http://dx.doi.org/10.1103/PhysRevB.83.144526}{Phys. Rev. B}, {\bf{83}}, 144526 (2011).
\bibitem{Faucher} M. Faucher, T. Fournier, B. Pannetier, C. Thirion, W. Wernsdorfer, J. C. Villegier, and V. Bouchiat,  \href{http://dx.doi.org/10.1016/S0921-4534(01)01168-6}{Physica C}, {\bf{368}}, 211 (2002).
\bibitem{Podd} G. J. Podd, G. D. Hutchinson, D. A. Williams, and D. G. Hasko,  \href{http://dx.doi.org/10.1103/PhysRevB.75.134501}{Phys. Rev. B}, {\bf{75}}, 134501 (2007).
\bibitem{Khapaev} M. M. Khapaev Jr, \href{http://dx.doi.org/10.1088/0953-2048/10/6/002}{Supercond. Sci. Technol.}, {\bf{10}}, 389 (1997).
\bibitem{Jesper} J. Johansson, K. Cedergren, T. Bauch, and F. Lombardi, \href{http://dx.doi.org/10.1103/PhysRevB.79.214513}{Phys. Rev. B}, {\bf{79}}, 214513 (2009).
\bibitem{Hasselbach} K. Hasselbach, D. Mailly,  and J. R. Kirtley,  \href{http://dx.doi.org/10.1063/1.1448864}{J. Appl. Phys.}, {\bf{91}}, 4432 (2002).
\bibitem{Tinkham} M. Tinkham, {Introduction to Superconductivity}, 2nd ed., McGrahaw-Hill International Editions, New York (1996).
\bibitem{Zaitsev} A. G. Zaitsev, R. Schneider, G. Linker, F. Ratzel, R. Smithey, P. Schweiss, J. Geerk, R. Schwab, and R. Heidinger, \href{http://dx.doi.org/10.1063/1.1435842}{Rev. Sci. Instrum.}, {\bf{73}}, 335 (2002).
\bibitem{Rosenthal} P. A. Rosenthal, M. R. Beasley, K. Char, M. S. Colclough, and G. Zaharchuk,  \href{http://dx.doi.org/10.1063/1.105660}{Appl. Phys. Lett.}, {\bf{59}}, 3482 (1991).
\bibitem{ClemPRB} J. R. Clem and E. H. Brandt, \href{http://dx.doi.org/10.1103/PhysRevB.72.174511}{Phys. Rev. B}, {\bf{72}}, 174511 (2005).
\bibitem{Sampietro} M. Sampietro, L. Fasoli, and G. Ferrari, \href{http://dx.doi.org/10.1063/1.1149785}{Rev. Sci. Instrum.}, {\bf{70}}, 2520 (1999).
\bibitem{Schwarz} T. Schwarz, J. Nagel, R. W\"{o}lbing, M. Kemmler, R. Kleiner, and D. Koelle, \href{http://dx.doi.org/10.1021/nn305431c}{ACS Nano}, {\bf{7}}, 844 (2013).
\bibitem{Nagel} J. Nagel, K. B. Konovalenko, M. Kemmler, M. Turad, R. Werner, E. Kleisz, S. Menzel, R. Klingeler, B. B\"{u}chner, R. Kleiner, and D. Koelle, \href{http://dx.doi.org/10.1088/0953-2048/24/1/015015}{Supercond. Sci. Technol.}, {\bf{24}}, 015015 (2011).
\bibitem{Drung} D. Drung, \href{http://dx.doi.org/10.1088/0953-2048/16/12/002}{Supercond. Sci. Technol.}, {\bf{16}}, 1320 (2003).

\end{thebibliography}

\end{document}